\documentstyle[epsf]{l-aa}

\newcommand{\re}{\mbox{$R_{E2}$}}
\newcommand{\be}{\mbox{$B_{E2}$}}

\begin{document}
\thesaurus{08.15.1,08.22.1,11.13.1}
\title{ EROS VARIABLE STARS  : Discovery of Beat Cepheids in the Small Magellanic Cloud and the
effect of metallicity on pulsation.} 
\author{ J.P. Beaulieu\inst{1,2}, M. Krockenberger$^{3}$, D.D. Sasselov$^{3}$, C. Renault$^{4}$,
R. Ferlet\inst{1}, A. Vidal-Madjar\inst{1}, E. Maurice\inst{5}, L. Pr\'evot\inst{5}, 
E. Aubourg$^{4}$, P. Bareyre\inst{4}, S. Brehin\inst{4}, C. Coutures\inst{4}, 
N. Delabrouille\inst{4}, J. de Kat\inst{4}, M. Gros\inst{4}, B. Laurent\inst{4}, 
M. Lachi\`eze-Rey\inst{4}, E. Lesquoy\inst{4}, C. Magneville\inst{4}, A. Milsztajn\inst{4}, 
L. Moscoso\inst{4}, F. Queinnec\inst{4}, 
J. Rich\inst{4} ,M. Spiro\inst{4}, L. Vigroux\inst{4}, S. Zylberajch\inst{4}, 
R. Ansari\inst{6}, F. Cavalier\inst{6}, M. Moniez\inst{6}, C. Gry\inst{7}. The EROS Collaboration \\ }
\institute{
Kapteyn Laboratorium, Postbus 800 9700 AV Groningen, The Netherlands
\and
Institut d'Astrophysique de Paris, CNRS, 98bis Boulevard. Arago, 75014 Paris, France
\and
Harvard-Smithsonian Center for Astrophysics, 60 Garden St., Cambridge, MA 02138, USA. 
\and
CEA, DSM/DAPNIA, Centre d'\'etudes de Saclay, 91191 Gif-sur-Yvette, France.
\and
Observatoire de Marseille, 2 place Le Verrier, 13248 Marseille 04, France.
\and
Laboratoire de l'Acc\'el\'erateur Lin\'eaire IN2P3, Centre d'Orsay, 91405 Orsay, France.
\and
Laboratoire d'Astronomie Spatiale CNRS, Travers\'ee du siphon, les trois lucs, 13120 Marseille, France.
 }
\offprints{JP. Beaulieu, beaulieu@astro.rug.nl}

\date{Received;accepted}
\maketitle
\markboth{J.P.\ Beaulieu et al.: Beat Cepheids in the SMC}{J.P.\ Beaulieu et al. : Beat Cepheids in the SMC}

\begin{abstract}
We report the discovery of eleven beat Cepheids in the Small Magellanic Cloud,
using data obtained by the EROS microlensing survey. Four stars are beating in the fundamental and
first overtone mode (F/1OT), seven are beating in the first and second overtone (1OT/2OT).
The SMC F/1OT ratio is systematically higher than the LMC F/1OT, while
the 1OT/2OT period ratio in the SMC Cepheids is the same as the LMC one. 

\end{abstract}

\section{ Introduction}

``Beat Cepheids'' (BC hereafter) are a rare phenomenon among classical Cepheids, 
because of the simultaneous presence of two radial modes of pulsations. They provide a good basis 
for the understanding  of Cepheid envelopes.
In our Galaxy, only 14 BCs are known (Pardo \& Poretti 1996, and reference therein): 13 pulsating 
in the fundamental (F) and first overtone (1OT) mode, and one (CO Aur) pulsating in the first and second 
overtone (2OT), out of a sample of about 200 objects which have sufficiently extensive photometry 
to detect or to rule out double-mode pulsation. Usually  data has been acquired in different 
conditions and bands,  therefore non-homogeneous datasets have to be used to make the Fourier analysis to show a clear
beating of two modes. Therefore, this kind of studies are very difficult.

Over the past five years, a great effort has been made to obtain light curves of millions of stars in the Magellanic
Clouds (MC) in the quest for dark matter through microlensing effects. Two enormous databases have been built,
one from the MACHO team (Welch et al. 1996 and references therein, W96 hereafter) , 
the other one from the EROS team (Renault et al. 1996,  Beaulieu \& Sasselov, 1996 and references therein).
In contrast to Galactic work, the study of variable stars in the MCs has many advantages : in a given 
galaxy, stars can be considered to be at the same distance, the foreground reddening is low and the differential
reddening is small. Therefore, the interpretation of observed differences is relatively straightforward. 
Over the course of the past few years, both teams have obtained photometry in both clouds, the Large Magellanic 
Cloud (LMC) and the Small Magellanic Cloud (SMC). 

In the LMC, Alcock et al. (1995), ( Al95 hereafter), and W96 reported the discovery of 73 
BCs from a sample of 1466 Cepheids:
29 are F/1OT, and 44 are 1OT/2OT. The F/1OT stars have a period ratio systematically greater by 0.01
than the Galactic ones at the same period. The Galactic star CO Aur has a period ratio very similar to the LMC
1OT/2OT. Double mode excitation is seen 20 \% of the time in LMC Cepheids with fundamental period smaller 
than 2.5 days, and 1OT/2OT modes are selected for period $P_{1OT}$ shorter than 1.25 days. The light curve geometry 
of each mode can be extracted (W96, Pardo \& Poretti 1996). 
% The morphological properties of each mode are very similar to single-mode pulsating Cepheids.

In this letter, we report the discovery of 11 BCs in the SMC from the EROS database. We will present a comparison
of their properties with the ones from the LMC and our galaxy in terms of period ratios and Fourier coefficients.

\section{ EROS Observations}

EROS is a  French collaboration
between astronomers and particle physicists to search for baryonic dark
matter in the form of compact objects in the Galactic Halo through 
microlensing effects on stars in the Magellanic Clouds.
Observations were done at ESO La Silla with a 0.4m f/10 reflecting telescope and a 16 CCD
camera in the 1993-1995 campaign (Arnaud et al., 1994a, Arnaud et al., 199b). 
 The SMC data set covering a field of $0.4 \times 1$ 
square degree is made of two colors light curves 
for $1.3 \times 10^5$ stars with excellent phase coverage ($\sim 3000$ points 
per bandpass). High-accuracy Fourier component are obtained.
 
Two broad band filters \be and \re are used.  
The \be filter is closer to Johnson V than to Johnson B, and is 
 broader than Johnson V. The \re bandpass is intermediate between Cousin R and I.
We defined the natural EROS color magnitude system as : a zero color star (a main sequence A0 star), 
\be - \re $\approx 0$
will have its \be magnitude numerically equal to its Johnson $V_J$
magnitude, and its \re   magnitude numerically equal to its Cousin
$R_C$ magnitude. 

\section{Beat Cepheids}
\subsection {Identification and Fourier decomposition}

Variable stars were identified and periods were determined simultaneously
using the One Way Analysis of Variance
(Schwarzenberg-Czerny  1989). This phase dispersion minimization method 
is based on strong statistical tests and is powerful to search for 
periodic signals of arbitrary shapes. For each trial period tested, a 
confidence level is yielded. 
We performed technical cuts in order to remove unreliable data and values
from the photometric time series. We excluded about 20 \% of the stars, and
20 \% of the measurements of a given star. Then we built AoV periodograms for
all the stars to search for periodicities in the range $0.1-100$ days. We 
excluded spurious periods due to chance fluctuation by the calculation of a 
probability of false detection. We kept all significant fluctuations in this
period range. We delineated the Cepheid instability strip in the color magnitude 
diagram to select Cepheid candidates. We plotted the period-luminosity relation
and excluded Population II Cepheids that are known to be about $\sim 1.5$ mag dimmer than 
Population I Cepheids. Then we performed a visual inspection of the light curves 
to exclude eclipsing binaries. Cepheids of period longer than $\sim 30-40$ days are
saturated on the detector, and therefore are not reconstructed. The remaining 450 
stars form our sample of SMC Cepheids. We adopted a Fourier decomposition of the form 
$ X = X_0 + \sum_{i=1}^M X_i \cos (i~\omega_i t + \Phi_i)$.
The customary quantities $R_{k1}={X_k / X_1}~k>1$ and 
$\Phi_{k1}=\Phi_k - k ~\Phi_1~k>1$  were defined. The amplitude ratio 
$R_{k1}$ reflect the asymmetry of the variation, and $\Phi_{k1}$ the 
full width at half maximum of the curve. 
We adopted the morphological classification proposed by Antonello et al., (1986),
and classify as s-Cepheids the stars that lie in the lower part of the $R_{21}-P$ plane,
and as classical Cepheids the remaining stars. This morphological separation is mirrored
by a clear dichotomy in the PL plane ( Beaulieu et al. 1995, Beaulieu \& Sasselov 1996)
showing again that s-Cepheids indeed pulsate in a different mode than classical Cepheids : 
first overtone for s-Cepheids and fundamental mode for classical Cepheids.
Stars strongly affected by blending effects have been marked, thanks to the examination of the loops
along one pulsational cycle in the luminosity temperature plane. They will be excluded from
further analysis.

 For each star, we compute the Fourier model corresponding to the most significant period, then do a sigma 
clipping at $1 \sigma$, then recompute the Fourier model. We substract it from the light
 curve and 
compute an AoV periodogram. The inspection of this periodogram and of the periodogram of the complete 
light curve allows us to select stars 
for which there is power in the spectrum due to another mode of pulsation. For beat Cepheids, 
power is expected in the spectrum at the frequencies of the two modes, plus the combination
of the harmonics.
This procedure is not the most efficient that could have been applied. 
However, it is a good way to select  BC candidates  and investigate their properties. 
A more detailed study, combining the two color light curves and involving improved signal processing 
techniques  like the CLEAN algorithm will have to be done in the future.

We found 11 BCs in our data base. It is not a complete sample, only the obvious BCs
have been discovered so far. 
Their magnitudes, colors, and period ratios are summarized in Table 1.
Four are pulsating in the the fundamental and first overtone (F/1OT) mode, seven are pulsating in the
first and second overtone mode (1OT/2OT). One can note that stars with period less than $\sim 1$day
are pulsating in 1OT/2OT, whereas stars with period greater than $\sim 1$day are pulsating in F/1OT.
Figure 1 shows luminosity as a function of period for SMC Cepheids with periods shorter than 4.5 days.
1OT/2OT lie above the first-overtone period-luminosity relation. 

\begin{figure}
\epsfxsize=9cm
\epsfysize=9cm
\centerline{\epsffile{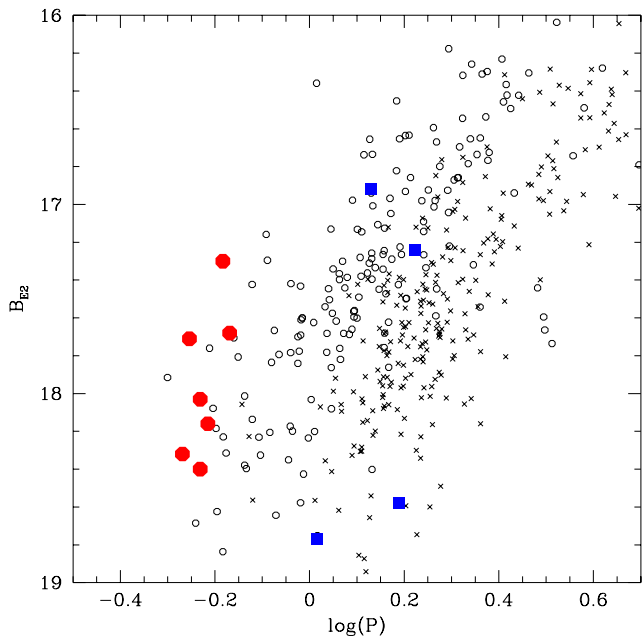}}
\caption[]{Luminosity as a function of period for SMC Cepheids from the EROS database. Classical
Cepheids (fundamental mode pulsators) are plotted as crosses, and s-Cepheids (overtone pulsators)
are plotted as open circles. Beat Cepheids are plotted as big dots (1OT/2OT) or as squares (F/1OT). 
For each star the adopted period is the one from the higher order mode.}
\end{figure}
%
%%  \begin{figure}
%% \epsfxsize=8cm
%% \epsfysize=8cm
%% \centerline{\epsffile{beat.cm.ps}}
%% \caption[]{The color magnitude diagram for short period  SMC Cepheids from the EROS database. Classical
%% Cepheids (fundamental mode pulsators) are plotted as crosses, and s-Cepheids (overtone pulsators)
%% are plotted as opened circles. Beat Cepheids are plotted as big dots (1OT/2OT) or as squared (F/1OT)  }
%% \end{figure}
%
%
\subsection{Period ratios}
\begin{figure}
\epsfxsize=9cm
\epsfysize=9cm
\centerline{\epsffile{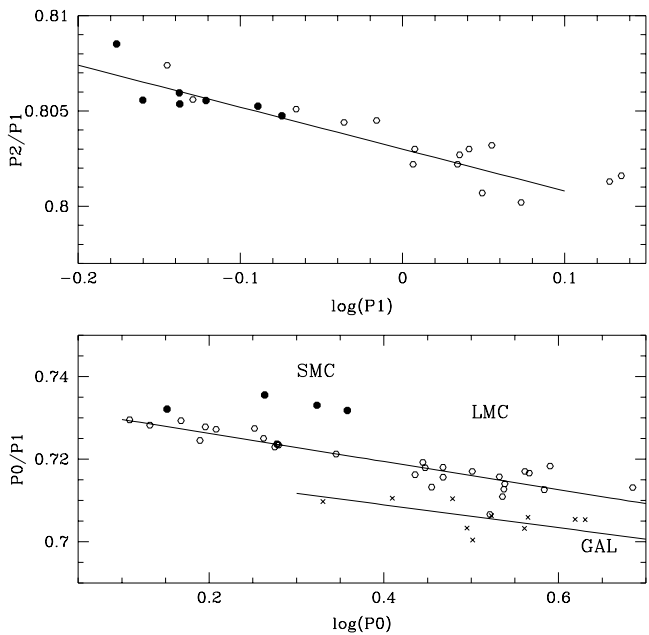}}
\caption[]{ Period ratio versus period for Galactic (crosses), LMC ( circle) and SMC (dots) Beat Cepheids. 
The linear fits come from Al95.}
\end{figure}

It is very striking to see in Figure 2 that the SMC 1OT/2OT pulsators have period ratios 
very similar to the LMC ones.
They lie on the linear fit to LMC 1OT/2OT ratios from Al95. 
The SMC F/1OT are offset from the LMC F/1OT relation by roughly the same amount as the LMC BCs from the Galactic BCs.
The heavy metals content of the Cepheids in the Galaxy, LMC and SMC are approximately 
0.016, 0.008 and 0.004 by mass, respectively. 

 From a theoretical point of view, by going to lower metallicities
there will be changes of the structure of the star : at fixed mass, the luminosity will
increase, therefore, the period ratio will decrease. When going to lower Z, the opacity bump that affects the 
pulsation will be smaller and therefore will increase the period ratio for a fixed mass and luminosity. 
The final result, the observed period ratio, will be a combination of these two antagonistic effects
plus an unknown non-linear coupling  effect.
In the case of a F/1OT pulsator, the fundamental mode penetrates deeper in the envelope of the star than the
first overtone. Therefore, the changed period ratio is mainly due to the effect of opacity on the fundamental
period. The case of 1OT/2OT is different since these two modes are more concentrated to the surface.
 We can crudely consider that an upward change in metallicity is an increase
of the opacity in the opacity bump due to metals found at depths around
100 000 K. The driving of the pulsation comes from helium ionization zone, while the metals opacity bump 
affects the acoustic properties of the envelope.
  
Therefore it has a much bigger effect on fundamental modes, than on higher-order overtone modes.
This can be seen  from linear adiabatic calculations (Christensen-Dalsgaard and Petersen 1995),
and from linear non-adiabatic calculations Buchler et al., 1996, Morgan and Welch 1996). 
In the case of F/1OT pulsators, the effect is bigger than the prediction by Morgan \& Welch (1996), 
but in the theoretically expected direction.
However, a  small shift was expected in the case of 1OT/2OT, and we see no evidence for it.

\section{Fourier coefficients}

We tried to extract the Fourier coefficients in order to characterize the morphological properties of the different
modes of pulsation. In most of the cases, the pulsation is strongly dominated by one mode for which we can easily derive
the two first orders of its  Fourier decomposition. However, the other mode has typically an amplitude 2 to 6 time smaller
(W96), which makes it difficult to isolate its properties without ambiguity. 
 We did a simultaneous fit of the two modes with three harmonics for the longer period mode, and two harmonics for the 
second mode of pulsation. The results are plotted in figure 3.
 
In the case of 1OT/2OT pulsators, we have detected the first overtone with amplitude X1 in the range 0.12-0.23 mag,
and the second overtone mode with amplitude X1 in the range 0.03-0.07 mag. 

In the case of F/1OT pulsators, the dominant mode is the first overtone. The fundamental mode has amplitude A1 in the range
0.05-0.09 mag, the first-overtone mode is in the range $0.14-0.22$ mag.
The second-overtone mode of pulsation has low $R_{21}$ values, and show a nice progression in the $\Phi_{21}-P$ plane.
It is striking that the mode of lowest amplitude of pulsation has a smaller $R_{21}$, which is not surprising 
as being more damped.

%cat Beat.Coeff.new | awk '{OFS=" & ";if ($12 != 0) print $1,$2,$3,$3/$12,$5,$6,$7,$8,$9,$12,$14,$15,$16,$17,$18}' |sort +4n
%cat tt | awk '{print $0,$11+0.68,$13+0.116}'
\begin{table}
%JPB NEW
\caption[]{SMC Beat Cepheids. CCD number and identification number in the EROS database, 
period of the longer mode of pulsation, period ratio, EROS colors and modes identification. }
\begin{flushleft}
\begin{tabular}{rllllllc}
\hline
ccd & iet & PerdL    &  Ratio  & $ B_{E2} $  & $R_{E2}$   & modes \\
\hline
3  & 9317 & 0.66622  &  0.8085 & 18.32 & 18.01 & 1OT/2OT \\
15 & 9594 & 0.69156  &  0.8056 & 17.71 & 17.14 & 1OT/2OT \\
13 & 2988 & 0.72833  &  0.8059 & 18.4  & 17.99 & 1OT/2OT \\
3  & 9830 & 0.72886  &  0.8054 & 18.03 & 17.70 & 1OT/2OT \\
11 & 3611 & 0.75643  &  0.8055 & 18.16 & 17.71 & 1OT/2OT \\
7  & 7614 & 0.81433  &  0.8052 & 17.3  & 16.71 & 1OT/2OT \\
3  & 1448 & 0.84246  &  0.8047 & 17.68 & 17.19 & 1OT/2OT \\
\hline 
9  & 2124 & 1.41844  &  0.7321 & 18.77 & 18.10 & F/1OT   \\
10 & 5185 & 2.10526  &  0.7330 & 18.58 & 17.75 & F/1OT   \\
10 & 3264 & 1.83486  &  0.7355 & 16.92 & 16.51 & F/1OT   \\
11 & 5498 & 2.28137  &  0.7318 & 17.24 & 16.68 & F/1OT   \\
\hline
\end{tabular}
\end{flushleft}
\end{table}
%plotted using four.sm macro (four2)  and four.beat.dat table

 \begin{figure}
 \epsfxsize=9cm
 \epsfysize=9cm
 \centerline{\epsffile{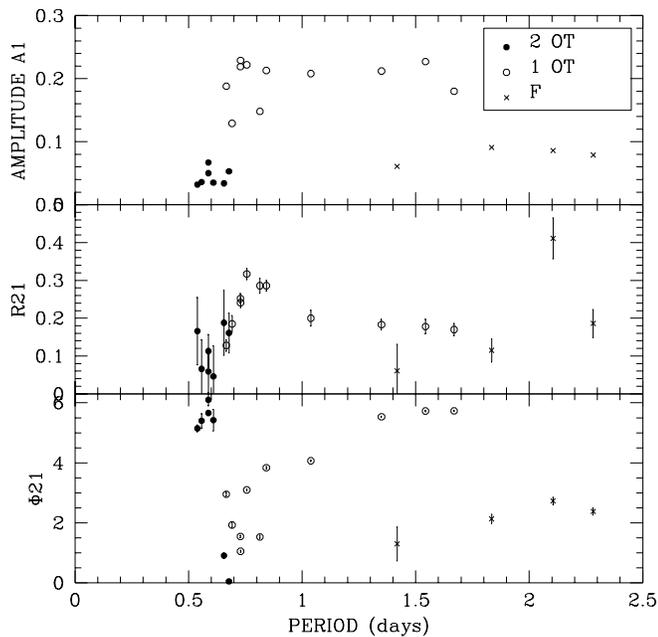}}
 \caption[]{The Fourier coefficients of SMC short period Cepheids from the EROS database. 
 Fundamental modes are plotted as crosses, first overtone modes
 are plotted as open circles, second overtone modes are plotted as filled circle. }
 \end{figure}

\section{Conclusion}

Eleven beat Cepheids were discovered as a by-product of the EROS microlensing survey in the Small Magellanic Cloud.
Four are pulsating in the fundamental and first overtone radial modes. Seven are pulsating in the
first and second overtone radial modes. This discovery was made thanks 
to the long term monitoring capability of EROS  and it underlines the power of this kind of database.

The SMC 1OT/2OTs are very similar to the LMC ones, while the F/1OTs have period ratios systematically
higher than in the LMC by $\sim$0.01. The amplitude of the first overtone is found to be in the range 0.12-0.23 mag, and 
the amplitude of the detected second overtones is in the range 0.03-0.07 mag. 

With these two kinds of beat Cepheids, observed at different metallicities, we are probing different depths in
the Cepheid envelopes, and drawing new strong constraints (similar to helioseismology) for the theory of stellar 
pulsation and the opacity tables at low metallicities. 

The light curves, Fourier models  and finding charts of these 11 Beat Cepheids are available on request to JP.Beaulieu 
(beaulieu@astro.rug.nl, beaulieu@iap.fr). An accurate astrometry is not yet available. 

\begin{acknowledgements}
This work is based on observations held at ESO La Silla. JPB gratefully acknowledges D. Welch 
for his pressure on us to obtain results on SMC Beat Cepheids, and his comments on a preliminary version of the article.
We thank J. Christensen-Dalsgaard for useful comments and the referee M. Feast.
\end{acknowledgements}

\end{document}